# Crystal growth and characterization of the antiperovskite superconductor MgC$_{1-x}$Ni$_{3-y}$

Nikolai D. Zhigadlo[a,b]

[a] *Department of Chemistry and Biochemistry, University of Bern, CH-3012 Bern, Switzerland*

[b] *CrysMat Company, CH-8046 Zurich, Switzerland*

**Abstract**

By varying the parameters controlling the growth of crystals, including the thermodynamic variables, such as temperature, pressure, and reagent composition and the kinetic factors, namely reaction time and cooling rate, we found the most appropriate conditions for the reproducible growth of the nonstoichiometric antiperovskite superconductors MgC$_{1-x}$Ni$_{3-y}$. Bulk single crystals of MgC$_{1-x}$Ni$_{3-y}$ were grown by a self-flux method at 3 GPa and 1700 °C using a mixture of Mg, C, and Ni powders in a molar ratio 1:1.25:3. The as-grown black colored crystals, mechanically extracted from solidified lump, exhibit various irregular three dimensional shapes, with flat surfaces and maximum dimensions up to ∼ (1-1.2) × (0.8-1.0) × (0.4-0.6) mm$^3$. Single-crystal x-ray diffraction refinement confirmed the high structural perfection of the grown crystals [Space group *Pm*-3*m*, No 221, *Z* = 1, *a* = 3.7913(1) Å, and V = 54.5(1) Å$^3$], but also the presence of deficiencies on the C and Ni sites. Temperature-dependent magnetization measurements showed a single-phase behaviour with a critical temperature ($T_c$) ranging between 6.3 and 6.8 K due to the slightly different C and Ni stoichiometries of MgC$_{1-x}$Ni$_{3-y}$ crystals. The growth of relatively large crystals reported here could provide a helpful guidance for further syntheses of various 3*d*-based antiperovskite intermetallics under high pressure.

Corresponding author.

*E-mail address*: nzhigadlo@gmail.com (N.D. Zhigadlo)

1. **Introduction**

The ABO$_3$ perovskite oxides are a well-known and a rich family of functional materials with many interesting physical properties [1]. There exists a related, less known family, antiperovskite oxides A$_3$BO (or BOA$_3$), in which the positions of the metal- and oxygen ions are reversed [2-4]. Furthermore, many 3$d$ transition-metal-based intermetallic non-oxide antiperovskites AXM$_3$, with A = Mg, Ca, Sc, Y, Lu, Zr, Nb, Ga, Al, Sn, Zn, Cu, In, Ge, etc.; X = B, C, N; and M = Mn, Fe, Ni, Ru, Rh, Pd, Pt, etc., have recently been studied. Such studies have revealed a range of intriguing physical properties and functionalities, including superconductivity [5-8], giant magnetoresistance [9], magnetostriction [10], negative thermal expansion [11], topological insulating property [12], and Dirac semimetal behaviour [13,14]. Based on theoretical predictions one can assume that M = Sc, Ti, V, Cr based antiperovskites may also show novel properties, such as superconductivity [15,16] and SbNMg$_3$, AsNBa$_3$, InNSc$_3$ could potentially act as improved thermoelectrics, superconductors, or topological insulators [17].

MgCNi$_3$ is the first reported classical nonoxide AXM$_3$-type antiperoskite superconductor, which provides a unique structural link between high-$T_c$ cuprate superconductors and intermetallic superconductors [5,18]. It has the typical three-dimensional CaTiO$_3$ cubic perovskite structure, but with the oxygen atoms on the faces here replaced by Ni atoms, and the Ca and Ti atoms replaced by Mg and C respectively; i.e., each C atom is surrounded by six Ni atoms forming hollow Ni$_6$-octahedra (Fig. 1). Despite the large number of experimental and theoretical works, the nature and microscopic origin of superconducting state is still under debate [19-22]. Many physical properties determined using polycrystalline samples show contradictory results [20]. For instance, the carbon content of MgC$_x$Ni$_3$ was reported to vary from below 10 to above 25 atomic per cent, thus making it difficult to assign a typical value to $x$ [23]. The single-phase perovskite structure of MgC$_x$Ni$_3$ can only be found in a narrow range of carbon content (0.88 < $x$ < 1.0) and the lattice constant ($a$) varies between 3.795 and 3.812 Å as $x$ varied from 0.887 to 0.978 [5,24]. Non-superconducting $\alpha$-MgC$_x$Ni$_3$ and superconducting $\beta$-MgC$_x$Ni$_3$ phases are found in the MgC$_x$Ni$_3$ system for $x < 1.0$ and $x > 1.0$ respectively, depending on the details of heat treatment [25]. An excess of C in the initial material is favourable to the formation of the superconducting phase [25,26]. Thus, despite very intense research, some puzzling questions remain concerning the dependence of superconducting properties on composition. More precise experiments using single-crystalline samples with well defined stoichiometry are indispensable for improving our understanding of superconductivity in MgCNi$_3$.

The main challenge in growing MgCNi$_3$ crystals is due to the huge differences in the melting temperatures of the starting components: 650 °C for Mg, 3827 °C for C, and 1455 °C for Ni [27]. Additionally, the high volatility of Mg and the relatively poor reactivity of C make it extremely difficult to synthesize the desired stoichiometry, i.e., Mg:C:Ni = 1:1:3, even in polycrystalline MgCNi$_3$. In order to compensate for the evaporated Mg and the less reactive C, excess of Mg and C should be supplied during the solid-state synthesis process. Because of these problems the synthesis of MgCNi$_3$ single crystals is considered as impossible in an open system at ambient pressure. However, it can be achieved by using high pressure in a closed system [21,27].

In order to grow sizable single crystals, suitable for various physical measurements, and to understand the effect of synthesis conditions and nonstoichiometry on the critical temperature ($T_c$), we carried out systematic investigations of the parameters controlling the growth of MgCNi$_3$. These included thermodynamic variables (temperature, pressure, and reagent composition), as well as kinetic factors, such as reaction time and cooling rate. We also discuss the possible crystallization mechanism, which accounts for the observed results. The successful growth of MgCNi$_3$ crystals reported here could also be extended to the synthesis of various AXM$_3$ antiperovskite analogues, thus opening up new possibilities for further exploration of these interesting materials.

## 2. Experimental details

For the growth of MgCNi$_3$ single crystals, we used the cubic-anvil, high-pressure, and high-temperature technique. The apparatus consist of a 1500-ton press, with a hydraulic-oil system comprising a semicylindrical multianvil module (Rockland Research Corp., USA). A set consists of eight large outer parts, which transmit the force through six small inner tungsten-carbide anvils (with edge length 22 mm) to the cubic high-pressure cell of ~27 × 27 × 27 mm$^3$ in a quasi isostatic way. More details about the design of the apparatus and the scheme of the high-pressure cell are reported in our previous publications [28, 29]. This method was successfully used earlier on to synthesize various superconducting and magnetic intermetallic crystals [30-32], diamonds [33], cuprate oxides [34,35], pyrochlores [36], *Ln*Fe*Pn*O (*Ln*: lanthanide, *Pn*: pnictogen) oxypnictides [37-42], and numerous other compounds [43-46].

A solid-state synthesized [47] MgCNi$_3$ powder precursor (abbreviated as MgCNi$_3$(P)) and pure elements of bright Mg flakes (99.998% Aldrich Chemicals), fine Ni powder (99.99% Johnson Matthey), and glassy graphite flakes (99.99% Alfa Aesar) were used as reactants. The starting mixtures of variable stoichiometries in total mass of 1 g were thoroughly mixed and pressed into pellets. The pellets were then placed in a cylindrical boron nitride (BN) crucible surrounded by a graphite-sleeve resistance heater and inserted into a pyrophyllite block which acted as a pressure medium. In a typical experiment, the assembled cell was compressed to 1 or 3 GPa at room temperature, and the compositions of starting mixtures, the heat-treatment temperature, the reaction time, and the cooling rate were tuned to determine the optimal conditions. After completing the crystal growth process, the pyrophyllite cube must be broken apart to get access to the BN crucible. Once the BN crucible was removed, the reaction products could easily be recovered, by cutting the crucible with a blade. No undesired reactions with the crucible were observed. In the optimized successful runs, the final product inside the crucible appeared as an almost fully-melted cylindrically-shaped solidified lump. This grown lump was mechanically crushed and the MgCNi$_3$ crystals were carefully extracted. The general morphology and dimensions of the crystals were determined by an optical microscope (Leica M205 C).

The X-ray single-crystal diffraction measurements were performed on an *Oxford Diffraction* Xcalibur PX area-detector diffractometer, which allowed us to examine the whole reciprocal space (Ewald sphere) for the presence of other phases or crystallites with different orientations. The crystal structure was determined by using a direct method and refined on $F^2$, employing the SHELXS-97 and SHELHL-97 programs [48]. The elemental analysis of the grown crystals was performed by energy dispersive X-ray spectroscopy (EDX). The temperature-dependent magnetization measurements were carried out using a Magnetic Property Measurement System (MPMS-XL, Quantum Design) equipped with a reciprocating-sample option.

### 3. Results and discussion

High-pressure and high-temperature (HPHT) methods are very effective not only in the search of new materials, but also for the growth of single crystals [49]. Nevertheless, due to a limited availability of high-pressure phase diagrams, it still remains an empirical technique based on rational trial and error. We note that the high-pressure technique used here for growing $MgCNi_3$ single crystals has several advantages: (i) the volatile Mg cannot escape from the high-pressure cell; (ii) at high temperatures Mg may act as a natural flux; (iii) at high-pressure and high-temperature the carbon reactivity becomes much higher; (iv) at high pressure all components are in close contact, which increases homogeneity.

We synthesized more than twenty samples with variable compositions and heat-treatment protocols. The most representative results are summarized in Table 1. The pressure was kept constant throughout the growth processes and was released only at the end of the crystal growth. The results of the different loadings and heat-treatment protocols can be summarized as follow: (i) The powdered precursor $MgCNi_3$(P), synthesized by solid-state reaction and heat-treated at pressure of 3 GPa by a protocols 1 and 2 result in polycrystalline samples with $T_c$ of 5.8 and 6.6 K, respectively. (ii) Heat-treatment of the mixture of $MgCNi_3$(P) with addition of 20 mol% of Mg (batches 3 and 4) also yields polycrystalline samples although slightly dense, most probably due to partial formation of the liquid phase. (iii) Further adding of 50 mol% of Mg to the $MgCNi_3$(P) precursor (batch 5) promote the formation of sufficient amount of liquid, however, the resulted products were not superconducting. (iv) The most successful results were obtained by using elemental Mg, C, and Ni powders as reactants for starting mixtures. In most cases significant amount of a liquid solidification are detectable in the reaction crucible after the experiments, thus supporting the assumption that MgCNi$_3$ crystallizes in a Mg liquid-state environment. Stoichiometric, i.e. 1:1:3 loading (batch 6) results in crystalline products with relatively high $T_c$ onset at ~ 7.3 K, however, with a broad transition. The best results show the C-excess loadings, i.e., 1:1.25:3 (batches 8 and 9), which led to single-crystalline samples with sharp superconducting transitions and with $T_c$ ranges between 6.3 and 6.8 K. The size of the grown crystals ranged from dozens micrometers to ~ (1-1.2) × (0.8-1.0) × (0.4-0.6) mm$^3$, as show in Fig. 2 (a,b). To further increase the size of crystals and achieve higher $T_c$ values, we attempted also slower cooling rates and annealing steps at 700 °C (batch 10). However, none of these attempts resulted in significant improvements.

The origin of the difference in $T_c$ between the various crystals is most likely due to slightly different sample stoichiometries [50,51]. Neutron diffraction results suggest that the highest $T_c$ of 7.3 K can be obtained for the stoichiometric composition MgCNi$_3$ [5]. Experiments performed on polycrystalline samples have shown that the superconductivity of MgC$_x$Ni$_3$ is very sensitive to the content of C and it disappears below about $x$ = 0.88. Thus, a C excess has to be added to the initial mixture to ensure attainment of superconducting composition. In our experimental conditions adding an excess of C = 1.25 seemed to favour the growth of superconducting crystals with a sharp transition. However, $T_c$ was surprisingly reduced, thus suggesting that the grown crystals had C deficiencies.

Let us now focus on the structural properties of the grown crystals. To check the structure of our crystals we collected diffraction data on several MgCNi$_3$ crystals, originating from 8 and 9 growth batches. For each case, a useful single crystals could be found and the refined structural model was essentially equivalent to the reported one [5]. Figure 2(c) presents a typical $hk$0 reciprocal lattice image for the MgCNi$_3$ single crystal with well-defined simple cubic structure. No additional phases, impurities, or intergrowing crystals were detected.

Figure 1 shows a 3D view of the crystal structure of MgCNi$_3$ with lattice parameter a = 3.7913(1) Å. In this structure there are six Ni atoms at the each cell forming a three-dimensional network of Ni octahedron. Each C atom is located at the body-centered cubic position surrounded by Ni octahedron cage. The Wyckoff positions for the atoms are Mg: $1a$ (0,0,0); C: $1b$ (0.5,0.5,0.5); Ni: $3c$ (0,0.5,0.5). A full X-ray refinement was performed on a crystal with dimensions 0.353 × 0.146 × 0.356 mm³, using 768 reflections (of which, 55 unique) in the κ-space region $-6 \leq h \leq 6$, $-6 \leq k \leq 6$, $-6 \leq l \leq 6$, by a full-matrix least-squares minimization of $F^2$. The weighting scheme was based on counting statistics and included a factor to down-weight the most intense reflections. After several refinement cycles, the correct crystallographic composition was determined and the final $R$ factor was 1.65%, indicating the high quality of the structural model. The details of the crystallographic parameters are summarized in Table 2. The refined atomic coordinates, the full thermal displacement parameters $U^{ij}$, the bond lengths and angles for all atoms are presented in Table 3 and 4. Since previous results suggest a lack of pure stoichiometry, we allowed the occupancies of the Mg, C, and Ni sites to vary in the refinement. The occupation parameters for the Mg, C, and Ni were found to range as follows: 1.00 ± 0.01, 0.92 ± 0.02, 2.88 ± 0.03, making the exact stoichiometry $Mg_{1.00 \pm 0.01}C_{0.92 \pm 0.02}Ni_{2.88 \pm 0.03}$. The Mg, C, and Ni contents determined by EDX are in a quantitative agreement (within ~ 5%) with X-ray refinement. Thus, despite the fact that in a high-pressure environment the reaction is drastically enhanced, surprisingly, we observed C and Ni vacancies in the final products. The small magnitudes of the isotropic displacement parameters obtained in the refinement indicate that C and Ni vacant positions did not introduce significant structural disorder. Lee et all. [27] also note that MgCNi$_3$ crystals grown at pressure of 4.25 GPa had Ni deficiencies, i.e. MgCNi$_{2.8}$ ($a$ = 3.812 Å, $T_c$ = 6.7 K) and those below 3.5 GPa had C deficiencies. Considered together these data reveal that besides the starting chemical composition and reaction temperature, pressure may be considered as another axis spanning the space in which certain stoichiometries may exist.

Figure 3 shows temperature dependence of dc magnetic susceptibility for MgCNi$_3$ samples synthesized with different starting materials and high-pressure and high-temperature conditions (see Table 1). For each successful batch (8, 9 and 10), several crystals were characterized by magnetic susceptibility measurements and were found to have a superconducting transition temperature between 6.3 and 6.8 K (Fig. 3 and 4). These values are slightly lower than the highest $T_c$ observed for polycrystalline MgCNi$_3$. This implies that C and Ni vacancies have a negative effect on the superconductivity and reduce the $T_c$ in MgC$_{1-x}$Ni$_{3-y}$ single crystals.

Interestingly, being deficient on both C and Ni sites the transition to the superconducting state for our single-crystalline samples is rather narrow ~0.35 K (Fig. 4). The good quality of the grown crystals was also confirmed by the resistivity studies reported in Ref. [21]. The first critical field, $H_{c1}$, determined from low-field magnetization curves is about 18 mT. The temperature dependence of the upper critical field, $H_{c2}$, is found to be isotropic with a slope at $T_c$ of -2.63 T/K and $H_{c2}(0) \approx 12.3$ T at zero temperature [21]. Both these $H_{c1}$ and $H_{c2}$ values are close to those observed for stoichiometric polycrystalline MgCNi$_3$ [20].

Figure 5 summarizes the variation of superconducting transition temperature $T_c$ with actual C ($x$) and Ni ($y$) content in polycrystalline MgC$_{1-x}$Ni$_3$ and single crystalline MgC$_{1-x}$Ni$_{3-y}$ samples. It is evident that the single crystalline samples where Ni deficiency is present demonstrate less sensitivity of their $T_c$ to the concentration of C vacancies. The occurrence of C and Ni vacancies may influence $T_c$ through a modification in the value of the density of states at Fermi energy $N(E_F)$. Theoretically, the superconductivity originates from the Ni 3$d$ band, and the C $p$ orbital strongly hybridizes with the Ni 3$d$ orbital [14]. In order to understand the effects of this kind of defects the antiperovskite superconductor crystal with particular stoichiometry MgC$_{0.93}$Ni$_{2.85}$ ($T_c$ ~ 6.5 K) was studied using high-resolution x-ray Compton scattering combined with electronic structure calculations [22]. These studies indicate that the Fermi surface is smeared by the disorder due to the presence of vacancies on the C and Ni sites, but does not drastically change its shape. The reduction in $N(E_F)$ may explain why single crystals with C and Ni vacancies are observed to have a lower $T_c$ than stoichiometric polycrystals. Interestingly, the transition temperature of MgC$_{0.93}$Ni$_{2.85}$ is higher than the $T_c \approx 4.5$ K expected for polycrystals with the C$_{0.93}$ site occupation and a fully occupied Ni site [24]. Nevertheless, this finding needs further clarification since polycrystalline samples besides the main perovskite phase contains tiny peaks of MgO impurity.

The appearance of vacancies on C and Ni sites raises the question of whether the growth of perfectly stoichiometric samples is thermodynamically possible under high-pressure conditions. So far, only compounds having certain composition deficiencies and a reduced $T_c$ could be stabilized at HPHT conditions. To clarify this issue, DFT calculations at high-pressure and high-temperature condition are necessary. One should also consider that carbides often exhibit vacancies in their sublattices. For example, we note here that $Mo_3Al_2C$ samples show a rather wide spread of $T_c$ = 8.6-9.3 K, due to carbon deficiencies [31]. All together, these results suggest that the studied compound might be stabilized by vacancy formation. Further synthesis conditions should be tested to overcome the vacancies issue, thereby allowing the achievement of the higher $T_c$ values expected for stoichiometric $MgCNi_3$.

## 4. Conclusion

In summary, we could show that high-pressure synthesis is a promising way to grow sizable nonoxide antiperovskite $MgCNi_3$ single crystals. The starting mixtures made of Mg, C, and Ni elements in the $MgC_{1.25}Ni_3$ stoichiometry, heat treated at 1700 °C and a pressure of 3 GPa, resulted in superconducting bulk crystals. Single-crystal x-ray diffraction refinement confirmed the high structural perfection of the grown crystals with C- and Ni-deficient stoichiometries. Due to these deficiencies, a certain variability in critical temperature ($T_c$) between 6.3 and 6.8 K was observed. At the same time, the superconducting transition turned out to be very narrow, $\Delta T_c$ = 0.35 K. The current work may provide a helpful guidance to the synthesis of analogous antiperovskite intermetallics under high-pressure conditions, thus opening up new possibilities for the further exploration of these interesting materials.

### Acknowledgements

I would like to acknowledge T. Klimczuk for providing $MgCNi_3$ powder sample used as precursor in the crystal growth experiments. I also thank S. Katrych and D. Logvinovich for collaboration on the early stage of this study and T. Shiroka for critically reading the manuscript and helpful comments.

**Table 1.** Experimental details and results of MgCNi$_3$ samples synthesized with different starting materials and high-pressure and high-temperature conditions.

| Batch | Starting materials | Pressure | Heat-treatment protocols | Notes on the products |
|---|---|---|---|---|
| 1 | MgCNi$_3$(P) | 3 GPa | ↗1500 °C/1 h, →1500 °C/0.5 h, ↘700 °C/3 h, ↘RT/0.5 h | ceramic, $T_c$ = 5.8 K, SF ≈ 84 % |
| 2 | MgCNi$_3$(P) | 3 GPa | ↗1850 °C/1 h, →1850 °C/0.3 h, ↘1000 °C/2 h, ↘RT/0.5 h | ceramic, $T_c$ = 6.6 K, SF ≈ 89 % |
| 3 | MgCNi$_3$(P) + 0.2Mg | 3 GPa | ↗1200 °C/1 h, →1200 °C/1 h, ↘600 °C/3 h, ↘RT/1.5 h | dense ceramic, $T_c$ = 6.6 K, SF ≈ 82 % |
| 4 | MgCNi$_3$(P) + 0.2Mg | 1 GPa | ↗1200 °C/1 h, →1200 °C/0.5 h, ↘600 °C/3 h, ↘RT/1.5 h | partially melted, $T_c$ = 5.8 K, SF ≈ 83 % |
| 5 | MgCNi$_3$(P) + 0.5Mg | 3 GPa | ↗1200 °C/1 h, →1200 °C/1 h, ↘600 °C/3 h, ↘RT/1.5 h | melted, not superconducting |
| 6 | Mg + C + 3Ni | 3 GPa | ↗1700 °C/1 h, →1700 °C/0.2 h, ↘RT/1 h | melted, $T_c$ = 7.3 K (broad), SF ≈ 79 % |
| 7 | Mg + 1.25C + 3Ni | 1 GPa | ↗1700 °C/1 h, →1700 °C/1 h, ↘RT/3 h | melted, crystals, not superconducting |
| 8 | Mg + 1.25C + 3Ni | 3 GPa | ↗1700 °C/1.5 h, →1700 °C/0.5 h, ↘300 °C/10 h, ↘RT/1 h | melted, crystals, $T_c$ = 6.3 – 6.8 K, SF ≈ 100 % |
| 9 | Mg + 1.25C + 3Ni | 3 GPa | ↗1700 °C/1.5 h, →1700 °C/0.5 h, ↘300 °C/15 h, ↘RT/1 h | melted, crystals, $T_c$ = 6.3 – 6.8 K, SF ≈ 100 % |
| 10 | Mg + 1.25C + 3Ni | 3 GPa | ↗1700 °C/4 h, →1700 °C/0.5 h, ↘700 °C/50 h, →700 C/5 h, ↘RT/2 h | melted, crystals, $T_c$ = 6.3 – 6.8 K, SF ≈ 100 % |

Note: MgCNi$_3$(P) – powder precursor with $T_c$ = 6.6 K synthesized by solid-state reaction at ambient condition; symbols description: ↗ - increase temperature in x hour; → - dwell temperature in x hour; ↘ - decrease temperature in x hour; $T_c$ – superconducting critical temperature; for batches 1 – 7 $T_c$ was determined by measuring of 1 or 2 big size pieces; for batches 8 – 10 $T_c$ was determined by measuring of 5-7 small size pieces from the same process. The shielding fraction (SF) was estimated from the magnetic measurements at 5 K by the zero-field cooling method.

**Table 2.** Details of single-crystal X-ray diffraction data collection and crystal refinement results for $MgC_{0.92}Ni_{2.88}$.

| | |
|---|---|
| Identification code | shelx |
| Empirical formula | $MgC_{0.92}Ni_{2.88}$ |
| Formula weight | 188.79 g/mol |
| Temperature | 298(2) K |
| Wavelength | Mo $K_\alpha$ (0.71073 Å) |
| Crystal system | Cubic |
| Space group | *Pm*-3*m* |
| Unit cell dimensions | $a = b = c = 3.7913(1)$ Å,  $\alpha = \beta = \gamma = 90°$ |
| Cell volume | 54.496(2) Å$^3$ |
| Z | 1 |
| Density (calculated) | 5.753 Mg/m$^3$ |
| Absorption coefficient | 22.458 mm$^{-1}$ |
| F(000) | 91 |
| Crystal size | $0.353 \times 0.146 \times 0.356$ mm$^3$ |
| $\theta$ range for data collection | 5.38 - 40.50° |
| Index ranges | $-6 \leq h \leq 6, -6 \leq k \leq 6, -6 \leq l \leq 6$ |
| Reflections collected | 768 |
| Independent reflections | 55 [$R_{int}$ = 0.0556] |
| Completeness to $\theta$ = 40.50°, % | 100.0 % |
| Absorption correction | Gaussian |
| Refinement method | Full-matrix least-squares on $F^2$ |
| Data / restraints / parameters | 55 / 0 / 8 |
| Goodness-of-fit on $F^2$ | 1.307 |
| Final R indices [$I > 2\sigma(I)$] | $R_1$ = 0.0165, w$R_2$ = 0.0432 |
| R indices (all data) | $R_1$ = 0.0165, w$R_2$ = 0.0432 |
| Extinction coefficient | 0.81(17) |
| Largest diff. peak and hole | 1.013 and -0.372 e.Å$^{-3}$ |

**Table 3.** Atomic coordinates, anisotropic displacement parameters ($U^{ij}$) and equivalent ($U^{eq}$) isotropic displacement parameters (in Å) for $MgC_{0.92}Ni_{2.88}$. The anisotropic displacement factor exponent takes the form: $-2\pi^2[h^2 a^{*2} U^{11} + ... + 2hk a^* b^* U^{12}]$. $U$(eq) is defined as one third of the trace of the orthogonalized $U^{ij}$ tensor. The results were obtained at room temperature and ambient pressure with the space group *Pm*-3*m*.

| Atom | Wyckoff | x | y | z | $U^{11}$ | $U^{22}$ | $U^{33}$ | $U^{23}$ | $U^{13}$ | $U^{12}$ | $U^{eq}$ |
|---|---|---|---|---|---|---|---|---|---|---|---|
| Mg1 | 1*a* | 0 | 0 | 0 | 0.009(1) | 0.009(1) | 0.009(1) | 0 | 0 | 0 | 0.012(1) |
| Ni2 | 3*c* | 0 | 0.5 | 0.5 | 0.014(1) | 0.010(1) | 0.014(1) | 0 | 0 | 0 | 0.014(1) |
| C3 | 1*b* | 0.5 | 0.5 | 0.5 | 0.013(1) | 0.013(1) | 0.013(1) | 0 | 0 | 0 | 0.013(1) |

**Table 4**. Bond lengths (Å) and angles (°) for $MgC_{0.92}Ni_{2.88}$.

| | | | |
|---|---|---|---|
| Mg1-Ni2#1 | 2.6809(1) | Ni2-Ni2#13 | 2.6809(1) |
| Mg1-Ni2 | 2.6809(1) | Ni2-Ni2#11 | 2.6809(1) |
| Mg1-Ni2#2 | 2.6809(1) | Ni2-Mg1#14 | 2.6809(1) |
| Mg1-Ni2#3 | 2.6809(1) | Ni2-Ni2#8 | 2.6809(1) |
| Mg1-Ni2#4 | 2.6809(1) | Ni2-Ni2#15 | 2.6809(1) |
| Mg1-Ni2#5 | 2.6809(1) | Ni2-Mg1#16 | 2.6809(1) |
| Mg1-Ni2#6 | 2.6809(1) | Ni2-Ni2#17 | 2.6809(1) |
| Mg1-Ni2#7 | 2.6809(1) | Ni2-Ni2#4 | 2.6809(1) |
| Mg1-Ni2#8 | 2.6809(1) | Ni2-Ni2#18 | 2.6809(1) |
| Mg1-Ni2#9 | 2.6809(1) | C3-Ni2#13 | 1.8956(2) |
| Mg1-Ni2#10 | 2.6809(1) | C3-Ni2#19 | 1.8956(2) |
| Mg1-Ni2#11 | 2.6809(1) | C3-Ni2#4 | 1.8956(2) |
| Ni2-C3#12 | 1.8956(2) | C3-Ni2#15 | 1.8956(2) |
| | | | |
| Ni2#1-Mg1-Ni2 | 60.000 | Mg1-Ni2-Ni2#11 | 60.000 |
| Ni2#1-Mg1-Ni2#2 | 120.000 | Ni2#13-Ni2-Ni2#11 | 180.000 |
| Ni2-Mg1-Ni2#2 | 180.000 | C3#12-Ni2-Mg1#14 | 90.000 |
| Ni2#1-Mg1-Ni2#3 | 180.000 | C3-Ni2-Mg1#14 | 90.000 |
| Ni2-Mg1-Ni2#3 | 120.000 | Mg1-Ni2-Mg1#14 | 180.000 |
| Ni2#2-Mg1-Ni2#3 | 60.000 | Ni2#13-Ni2-Mg1#14 | 60.000 |
| Ni2#1-Mg1-Ni2#4 | 60.000 | Ni2#11-Ni2-Mg1#14 | 120.000 |
| Ni2-Mg1-Ni2#4 | 60.000 | C3#12-Ni2-Ni2#8 | 45.000 |
| Ni2#2-Mg1-Ni2#4 | 120.000 | C3-Ni2-Ni2#8 | 135.000 |
| Ni2#3-Mg1-Ni2#4 | 120.000 | Mg1-Ni2-Ni2#8 | 60.000 |
| Ni2#1-Mg1-Ni2#5 | 120.000 | Ni2#13-Ni2-Ni2#8 | 120.000 |
| Ni2#1-Mg1-Ni2#5 | 120.000 | Ni2#11-Ni2-Ni2#8 | 60.000 |
| Ni2#2-Mg1-Ni2#5 | 60.000 | Mg1#14-Ni2-Ni2#8 | 120.000 |
| Ni2#3-Mg1-Ni2#5 | 60.000 | C3#12-Ni2-Ni2#15 | 135.000 |
| Ni2#4-Mg1-Ni2#5 | 180.000 | C3-Ni2-Ni2#15 | 45.000 |
| Ni2#1-Mg1-Ni2#6 | 120.000 | Mg1-Ni2-Ni2#15 | 120.000 |
| Ni2-Mg1-Ni2#6 | 90.000 | Ni2#13-Ni2-Ni2#15 | 60.000 |
| Ni2#2-Mg1-Ni2#6 | 90.000 | Ni2#11-Ni2-Ni2#15 | 120.000 |
| Ni2#3-Mg1-Ni2#6 | 60.000 | Mg1#14-Ni2-Ni2#15 | 60.000 |
| Ni2#4-Mg1-Ni2#6 | 60.000 | Ni2#8-Ni2-Ni2#15 | 180.000 |
| Ni2#5-Mg1-Ni2#6 | 120.000 | C3#12-Ni2-Mg1#16 | 90.000 |
| Ni2#1-Mg1-Ni2#7 | 60.000 | C3-Ni2-Mg1#16 | 90.000 |
| Ni2-Mg1-Ni2#7 | 120.000 | Mg1-Ni2-Mg1#16 | 90.000 |
| Ni2#2-Mg1-Ni2#7 | 60.000 | Ni2#13-Ni2-Mg1#16 | 60.000 |
| Ni2#3-Mg1-Ni2#7 | 120.000 | Ni2#11-Ni2-Mg1#16 | 120.000 |
| Ni2#4-Mg1-Ni2#7 | 90.000 | Mg1#14-Ni2-Mg1#16 | 90.000 |
| Ni2#5-Mg1-Ni2#7 | 90.000 | Ni2#8-Ni2-Mg1#16 | 60.000 |
| Ni2#6-Mg1-Ni2#7 | 120.000 | Ni2#15-Ni2-Mg1#16 | 120.000 |
| Ni2#1-Mg1-Ni2#8 | 120.000 | C3#12-Ni2-Ni2#17 | 45.000 |
| Ni2-Mg1-Ni2#8 | 60.000 | C3-Ni2-Ni2#17 | 135.000 |
| Ni2#2-Mg1-Ni2#8 | 120.000 | Mg1-Ni2-Ni2#17 | 120.000 |
| Ni2#3-Mg1-Ni2#8 | 60.000 | Ni2#13-Ni2-Ni2#17 | 120.000 |
| Ni2#4-Mg1-Ni2#8 | 90.000 | Ni2#11-Ni2-Ni2#17 | 60.000 |
| Ni2#5-Mg1-Ni2#8 | 90.000 | Mg1#14-Ni2-Ni2#17 | 60.000 |
| Ni2#6-Mg1-Ni2#8 | 60.000 | Ni2#8-Ni2-Ni2#17 | 90.000 |
| Ni2#7-Mg1-Ni2#8 | 180.000 | Ni2#15-Ni2-Ni2#17 | 90.000 |
| Ni2#1-Mg1-Ni2#9 | 60.000 | Mg1#16-Ni2-Ni2#17 | 120.000 |
| Ni2-Mg1-Ni2#9 | 90.000 | C3#12-Ni2-Ni2#4 | 135.000 |
| Ni2#2-Mg1-Ni2#9 | 90.000 | C3-Ni2-Ni2#4 | 45.000 |
| Ni2#3-Mg1-Ni2#9 | 120.000 | Mg1-Ni2-Ni2#4 | 60.000 |
| Ni2#4-Mg1-Ni2#9 | 120.000 | Ni2#13-Ni2-Ni2#4 | 60.000 |
| Ni2#5-Mg1-Ni2#9 | 60.000 | Ni2#11-Ni2-Ni2#4 | 120.000 |
| Ni2#6-Mg1-Ni2#9 | 180.000 | Mg1#14-Ni2-Ni2#4 | 120.000 |
| Ni2#7-Mg1-Ni2#9 | 60.000 | Ni2#8-Ni2-Ni2#4 | 90.000 |
| Ni2#8-Mg1-Ni2#9 | 120.000 | Ni2#15-Ni2-Ni2#4 | 90.000 |
| Ni2#1-Mg1-Ni2#10 | 90.000 | Mg1#16-Ni2-Ni2#4 | 60.000 |
| Ni2-Mg1-Ni2#10 | 120.000 | Ni2#17-Ni2-Ni2#4 | 180.000 |
| Ni2#2-Mg1-Ni2#10 | 60.000 | C3#12-Ni2-Ni2#18 | 45.000 |
| Ni2#3-Mg1-Ni2#10 | 90.000 | C3-Ni2-Ni2#18 | 135.000 |
| Ni2#4-Mg1-Ni2#10 | 60.000 | Mg1-Ni2-Ni2#18 | 120.000 |
| Ni2#5-Mg1-Ni2#10 | 120.000 | Ni2#13-Ni2-Ni2#18 | 90.000 |
| Ni2#6-Mg1-Ni2#10 | 60.000 | Ni2#11-Ni2-Ni2#18 | 90.000 |
| Ni2#7-Mg1-Ni2#10 | 60.000 | Mg1#14-Ni2-Ni2#18 | 60.000 |
| Ni2#8-Mg1-Ni2#10 | 120.000 | Ni2#8-Ni2-Ni2#18 | 60.000 |
| Ni2#9-Mg1-Ni2#10 | 120.000 | Ni2#15-Ni2-Ni2#18 | 120.000 |
| Ni2#1-Mg1-Ni2#11 | 90.000 | Mg1#16-Ni2-Ni2#18 | 60.000 |
| Ni2-Mg1-Ni2#11 | 60.000 | Ni2#17-Ni2-Ni2#18 | 60.000 |
| Ni2#2-Mg1-Ni2#11 | 120.000 | Ni2#4-Ni2-Ni2#18 | 120.000 |
| Ni2#3-Mg1-Ni2#11 | 90.000 | Ni2#13-C3-Ni2 | 90.000 |
| Ni2#4-Mg1-Ni2#11 | 120.000 | Ni2#13-C3-Ni2#19 | 90.000 |
| Ni2#5-Mg1-Ni2#11 | 60.000 | Ni2-C3-Ni2#19 | 180.000 |
| Ni2#6-Mg1-Ni2#11 | 120.000 | Ni2#13-C3-Ni2#4 | 90.000 |
| Ni2#7-Mg1-Ni2#11 | 120.000 | Ni2-C3-Ni2#4 | 90.000 |
| Ni2#8-Mg1-Ni2#11 | 60.000 | Ni2#19-C3-Ni2#4 | 90.000 |
| Ni2#9-Mg1-Ni2#11 | 60.000 | Ni2#13-C3-Ni2#15 | 90.000 |
| Ni2#10-Mg1-Ni2#11 | 180.000 | Ni2-C3-Ni2#15 | 90.000 |
| C3#12-Ni2-C3 | 180.000 | Ni2#19-C3-Ni2#15 | 90.000 |
| C3#12-Ni2-Mg1 | 90.000 | Ni2#4-C3-Ni2#15 | 180.000 |
| C3-Ni2-Mg1 | 90.000 | Ni2#13-C3-Ni2#1 | 180.000 |
| C3#12-Ni2-Ni2#13 | 135.000 | Ni2-C3-Ni2#1 | 90.000 |
| C3-Ni2-Ni2#13 | 45.000 | Ni2#19-C3-Ni2#1 | 90.000 |
| Mg1-Ni2-Ni2#13 | 120.000 | Ni2#4-C3-Ni2#1 | 90.000 |
| C3#12-Ni2-Ni2#11 | 45.000 | Ni2#15-C3-Ni2#1 | 90.000 |
| C3-Ni2-Ni2#11 | 135.000 | | |

Symmetry transformations used to generate equivalent atoms: #1 x, y, z; #2 -1+x, y, -1+z; #3 y, -1+z, -1+x; #4 z, x, y; #5 -1 +z, -1+x, y; #6 x, y, -1+z; #7 -1+z, x, y; #8 z, -1+x, y; #9 -1+x, y, z; #10 y, z, -1+x; #11 y, -1+z, x; #12 x, -1+y, z; #13 1+y, z, x; #14 1+x, y, 1+z; #15 z, x, 1+y; #16 1+x, y, z; #17 z, -1+x, 1+y; #18 1+y, -1+z, x; #19 x, 1+y, z.

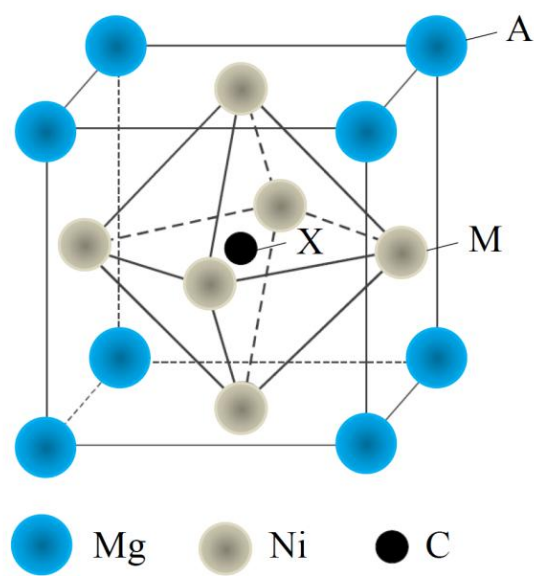

**Figure 1.** The antiperovskite AXM$_3$ type crystal structure of MgCNi$_3$. The coordinates of all three ions is reversed in such way that the A – sites are located at the corners of a cube, where the X is located in the centre of an M$_6$ octahedron. Typically, in this structure A is main group element; X = B, C, and N; and M is 3$d$ transition metal element.

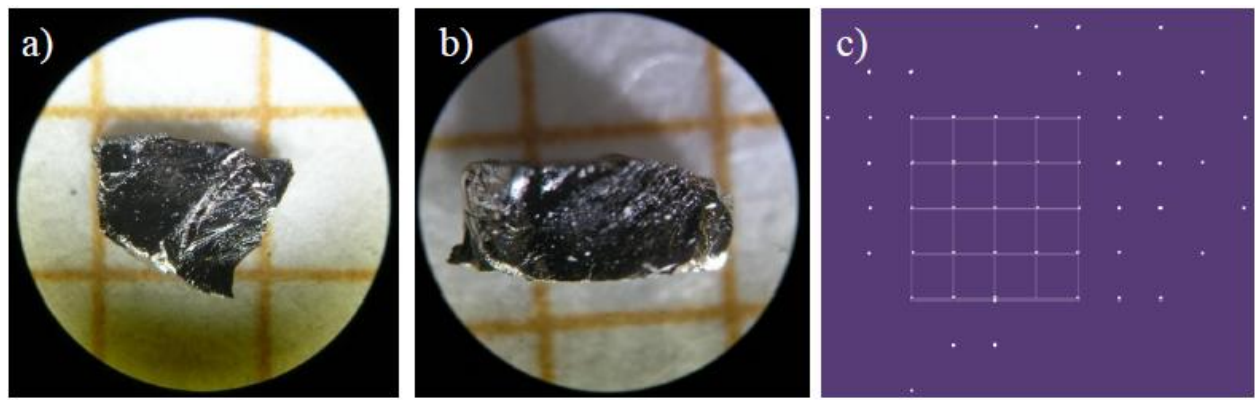

**Figure 2.** Optical microscope images of MgC$_{1-x}$Ni$_{3-y}$ crystals mechanically extracted from as-grown solidified lump. The edge of each individual picture is about 1 mm. Right frame shows the $hk0$ reciprocal space section determined by XRD for single-crystal MgC$_{0.92}$Ni$_{2.88}$. Well-defined simple cubic structure is seen.

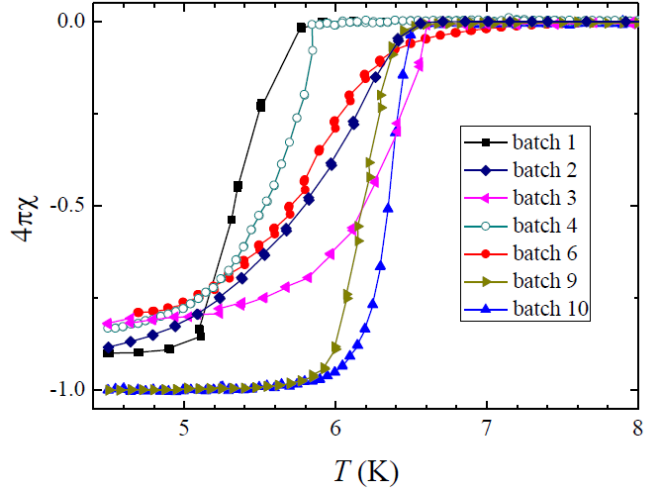

**Figure 3.** Temperature dependence of dc magnetic susceptibility for $MgCNi_3$ samples synthesized with different starting materials and high-pressure and high-temperature conditions (see Table 1). The zero-field-cooled curves were obtained on heating in a magnetic field of 10 Oe after zero-field cooling to 4.5 K.

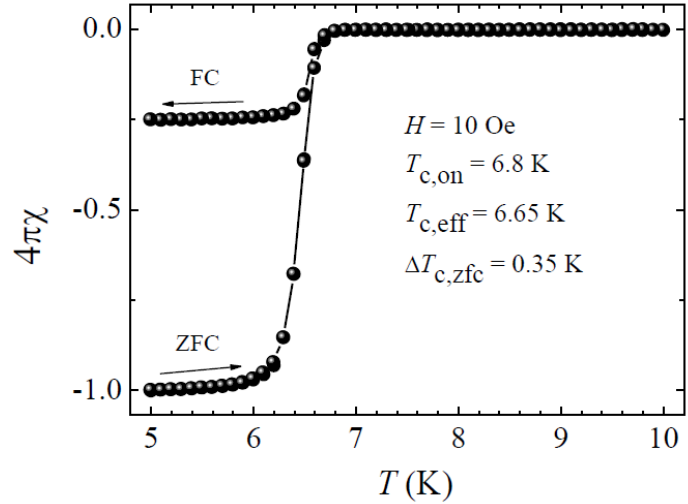

**Figure 4.** Temperature dependence of dc magnetic susceptibility for $MgC_{0.92}Ni_{2.88}$ crystal showing a superconducting transition at $T_c$ = 6.8 K (batch 8). The zero-field-cooled (ZFC) curve was obtained on heating in a magnetic field of 10 Oe after zero-field cooling to 5 K, and the field-cooled (FC) curve was done successively on cooling in the same field.

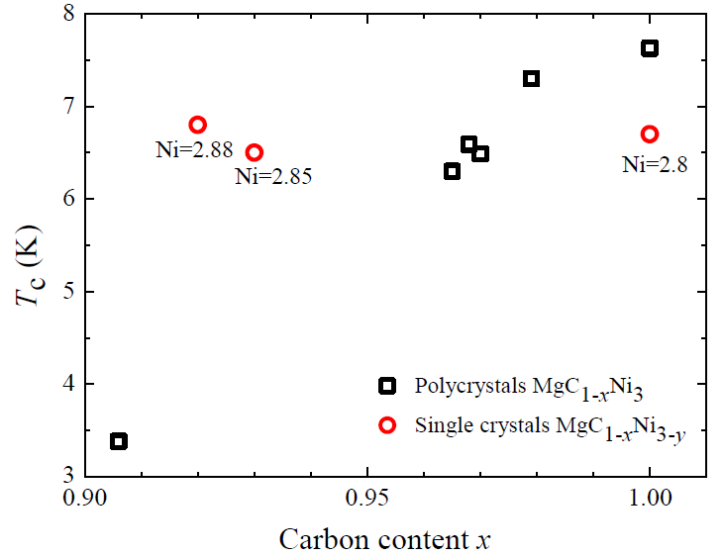

**Figure 5.** Variation of superconducting transition temperature $T_c$ with actual C ($x$) and Ni ($y$) content in polycrystalline $MgC_{1-x}Ni_3$ [24, 52] and single crystalline $MgC_{1-x}Ni_{3-y}$ [22, 27, current work] samples.